# Driver Identification by Neural Network on Extracted Statistical Features from Smartphone Data

**Ruhallah Ahmadian, Mehdi Ghatee[1]**

Department of Computer Science, Amirkabir University of Technology, Tehran, Iran.

**Abstract**

The future of transportation is driven by the use of artificial intelligence to improve living and transportation. This paper presents a neural network-based system for driver identification using data collected by a smartphone. This system identifies the driver automatically, reliably and in real-time without the need for facial recognition and also does not violate privacy. The system architecture consists of three modules data collection, preprocessing and identification. In the data collection module, the data of the accelerometer and gyroscope sensors are collected using a smartphone. The preprocessing module includes noise removal, data cleaning, and segmentation. In this module, lost values will be retrieved and data of stopped vehicle will be deleted. Finally, effective statistical properties are extracted from data-windows. In the identification module, machine learning algorithms are used to identify drivers' patterns. According to experiments, the best algorithm for driver identification is MLP with a maximum accuracy of 96%. This solution can be used in future transportation to develop driver-based insurance systems as well as the development of systems used to apply penalties and incentives.

## 1- Introduction

Artificial intelligence and data mining are two important strategies for the development of future transportation systems. Given the spread of traffic data, especially in the field of sensors, automobiles, IoT, telecommunications tools and smartphones, more attention is needed to these two strategies in transportation. Applications like connected vehicle control, traffic prediction according to big-data, and online decision-making in advanced driver-assistance systems and smart. The purpose of this article is to use artificial intelligence to identify a driver from a set of drivers based on smartphone data. This system allows us to identify the driver using driving characteristics such as physical or behavioral characteristics. One example implemented in this regard is seen in [1].

[1] URL: www.aut.ac.ir/ghatee
Email: ghatee@aut.ac.ir

Driver identification has a variety of applications, including intercity public transport, freight transportation, and quality driving control, as well as it can be used for special insurance policies, the imposition of fines, and in-service program[2]. For driving safety, it can help detect abnormal driving behaviors in driving evaluation systems, to detect dangerous driving or under the influence of alcohol and drugs and also to detect distraction. Driver identification is also used in intelligent transportation fleet management systems to control the amount of driving per day and avoid using an unauthorized driver. Also in the insurance industry, intelligence insurance requires driver identification and evaluation to get the price appropriate to the level of driving of car owners. For example, some companies consider high speed, midnight driving, and braking to be dangerous [3][4][5][6].

In Iran, since 2020, third party insurance will be driver-driven, so that the driver will be insured instead of the car. Factors affecting the cost of insurance include female or male, young or old, high-risk and low-risk drivers. Driver identification in the insurance industry is used to identify the unauthorized driver and the driver who has an accident. Driver identification is used in advanced driver-assistance systems to provide targeted services, for example by identifying a member of the family who is driving, it adjusts the vehicle or route settings to suit that individual [7][8][9]. Alarm systems can also be equipped with more advanced tools using machine learning [10]. Today, alarm systems use GPS as a monitoring tool. However, these systems do not automatically detect theft, but using driver identification can be equipped with automatic detection.

On the other hand, the use of IoT in driver identification is very important. It is projected that by the year 2020, 20 billion Internet of Things devices will be used in the world. These days, cars are equipped with a variety of sensors to provide advanced driver-assistance systems such as adaptive navigation control, auto park and automatic emergency brake. These sensors in vehicles lead to advances in transportation research. One of the most popular research topics is the study of driving style to identify driver behavior. A review of research into the study of driving style and the role of IoT is discussed in [11][12].

In the past, drivers were identified using simpler methods such as ID cards or notes, later more sophisticated methods such as fingerprint, face recognition as well as innovative methods such as finger vein pattern recognition was replaced [13][14][15]. However, it is easier and safer to identify a driver using a sensor because in previous systems it was possible to cheat. For example, a driver could be replaced with an unauthorized driver after moving, but the sensors reduce the possibility of fraud due to continuous evaluation. Also, since the sensors are automatic, the driver does not need any extra action. On the other hand, the use of voice and face recognition tools violates driver privacy, so the use of sensors that do not violate driver privacy is very important [3][4][5]. In this paper, a tentative basis for appropriate policy discussion is provided to balance tool and privacy in vehicle-data sharing scenarios. To this end, the potential for privacy breach in collecting vehicle sensor data is investigated [16].

Here are some of the major sensors used in research based on hidden sensors. Inertial Measurement Unit (IMU) includes accelerometer and gyroscope sensors. Smartphone or OBD-II device can be used to collect the data. The accelerometer sensor converts the mechanical acceleration into an electrical signal accordingly. Acceleration is the rate of change of velocity of an object concerning time. The accelerometer has single-axis and multi-axis models that can measure the size and orientation of the acceleration as a vector. Accelerometers sense the forces that cause acceleration (both static and dynamic). An example of a static force is the force of gravity And for dynamic forces, vibrations can be mentioned. A gyroscope is a device used for measuring or maintaining orientation and angular velocity. device consisting of a wheel or disc mounted so that it can spin rapidly about an axis which is itself free to alter in direction. The orientation of the axis is not affected by the tilting of the mounting. The behavior of a gyroscope reflects the stability of angular momentum properties (kinetic energy value and its orientation as a vector value). The change in orientation due to external torque is negligible. Because the external torque is minimized by holding the device in a loop, its orientation is almost constant. The main task of gyroscopes is to create a coordinate frame of reference, and accelerometers measure the moving acceleration along the axes, which can be inertia or another reference frame, such as a ground-based frame.

The following are classified investigations related to driver identification:

A: Data analysis of in-vehicle data recorders

Since the CAN bus become a standard in cars, many researchers have attempted to use their sensor data to identify and evaluate drivers. Nathanael et al [17] proposed maneuver-based driver identification. In this study, the maneuvers of acceleration and deceleration are collected and then identified by linear discriminant analysis to identify the drivers. The identification accuracy in this study was reported by up to 61%. In [16], driver identification was performed with the data of a car with 15 different passengers collected from the CAN bus. The research includes statistical, descriptive, and repetitive features, and the best algorithms, sensors, and features are reported. In [18], driver identification was performed using only turn maneuvering data. The researcher classifies the turn position into 12 classes. The model uses 12 CAN bus signals received from a single-vehicle by multiple drivers. The result is reported for classification with two classes of 79.6% and five classes of 50.1%. In [19], driver identification has been identified with impostor detection. Researchers have used ELM to detect impostor. The ELM inputs are eight variables derived from the CAN bus. Driver identification accuracy has been 80% and impostor detection with a single driver assumed to be 80% and multiple drivers assumed to be 50%. In [8], driver identification without feature extraction was performed using the CAN bus features obtained from 10 drivers and a 99% accuracy was reported. In [20], three different datasets [21][22][23] have been studied, as well as the minimum time required for network training and testing. In [24], CAN bus data analyzed by IVDR, a system for measuring driver performance and vehicle movement. By designing the parallel processing of machine learning methods have achieved an average accuracy of 88%.

B: Data analysis of side sensors

Since there are still many vehicles that are not compatible with the CAN bus standard, the use of more general sensors for driver identification is justified. In [25], driver identification was performed with GPS data collected from smartphones. In this research, the car's location features such as acceleration, angular velocity, sudden change of lane are extracted from GPS statistical features and the driver is identified using Random Forest. The experiment was conducted on 9 groups of drivers separately, each group consisting of 4 or 5 drivers and each group having the same time and location. The average accuracy for all groups was reported to be 82.3%. In [26], driver identification was performed only using accelerometer sensor data collected by smartphones. In this research, we apply the PCA feature extraction method on statistical variables obtained from acceleration data. PCA provides a template for each driver. The results show that the acceleration signal is capable of identifying the driver, although the pattern presented by the PCA is similar for some drivers. In [27], driver identification was performed using the histogram of the values obtained from the accelerometer sensor. In this study, after deleting the stop times from the data, segmentation and then use the histogram for each window. Before that, outliers due to sensor bugs are deleted according to the histograms. MLP algorithm used for identifying drivers. Parameters of the number of features and the percentage of window overlap are also considered. Accuracy with two hours of training data and 10 drivers reported 95%. In [28], driver identification was performed using driving data in a simulated environment by 10 drivers. The acceleration and steering features are used with the HMM method. Accuracy reported 85%. Table 1 and 2 provides further details of the methods.

According to the results of this table, in this article, the focus will be on extracting inertial sensors data from the smartphone. Appropriate features are extracted according to statistical models such as histograms, averages, etc. Then the classification is performed using learning models such as neural network, nearest neighbor, decision tree, random forest, etc. Drivers are identified through a dataset consisting of 10 drivers, 10 cars and approximately 7 hours and 30 minutes of driving for each driver.

*Table 1: Top driver identification studies*

| Ref | Property | Features/Sensors | Model | Result |
|---|---|---|---|---|
| [8] | Identification of 10 driver | Using CAN-bus: Air pressure, engine properties, fuel consumption, friction torque, oil temperature, steering | PCA for feature extraction and Decision Tree for classification | 99% accuracy |
| [16] | Identifying 15 drivers with the same car, driving length, and driving environment | Using CAN-bus: Brake pedal positions, max engine torque, steering wheel, lateral acceleration, fuel consumption, etc. | Support Vector Machine, Random Forest, Bayesian Network, Nearest Neighbor for classification | 83.77% accuracy with brake pedal and 93% accuracy by all sensors in parking and 100% accuracy by all sensors in the road |
| [17] | Identifying 14 drivers older than 70 years | Using CAN-bus: Location and speed | Linear Discriminant Analysis (LDA) | 34% accuracy by maneuvers detection, 30% accuracy by speed and 60.5% accuracy in total |

*Table 2: Top driver identification studies*

| Ref | Property | Features/Sensors | Model | Result |
|---|---|---|---|---|
| [18] | Driver identification from 10 car and 64 driver | Using CAN-bus: Steering wheel angle, steering velocity, steering acceleration, vehicle velocity, vehicle heading, engine RPM, gas pedal and brake pedal positions, forward and acceleration, lateral acceleration, torque and throttle position | Random Forest | 76.9% accuracy for 2-drivers and 50.1% accuracy for 5-drivers |
| [19] | Driver identification and impostor detection, among 11 drivers | Using CAN-bus: Gas pedal, brake pedal, laser scanner, acceleration, and rotation rate | Feedforward Neural Network | 80% accuracy for all groups, 90% accuracy for 2-3 persons, 80% accuracy for impostor detection with one main driver |
| [24] | Driver identification from one year of 217 families | Using CAN-bus: Location, and maneuvers | Decision Tree, Random Forest, Support Vector Machine and Learning Vector Quantization (LVQ) | ~88% accuracy |
| [25] | Identifying 38 drivers with smartphone data in two months of driving | Speed, location, acceleration | Random Forest | 82.3% accuracy for groups of 5-6 persons |
| [26] | Identification of 5 drivers by acceleration variance averages | Statistical features extracted from acceleration | PCA for driving patterns extraction and comparing with a reference pattern | 60%-100% accuracy by 4-7 variables of the covariance matrix |
| [27] | Identification using the accelerometer for 13 drivers in 10 months | Acceleration histogram | Feedforward Neural Networks | 95% accuracy |
| [28] | Understanding driver behavior. 20 drivers in a simulator | Acceleration and rotation of the steering | Hidden Markov M | 85% accuracy |

## 2- Proposed System

The purpose of this section is to provide a smartphone-based data system for driver identification. Figure 1 shows the overall architecture of the system. In this architecture, there are three modules, data collection, preprocessing and driver identification. This architecture is used separately in two phases of train and test. In the data collection module, assumed that the data comes from accelerometers and a gyroscope. In the preprocessing module, at first, the input signals are cleared and then reoriented, the smartphone may not be the same for all drivers in the same direction. The full description of reorientation is provided in [29]. The data cleaning steps are illustrated in Figure 2. As shown in this figure, in the data cleaning, the input signals are noise cleaned and then the samples containing the missing values are removed or completed. Since the times when the car is stopped for a long time, the data is not effective in analyzing the driver's behavior, we delete the data related to the car's stop time if the stop time is longer than 6 seconds. Aggregation of axes of acceleration was calculated to detect stop time, if the value is unchanged with a 0.5 threshold, meaning that the vehicle has not moved.

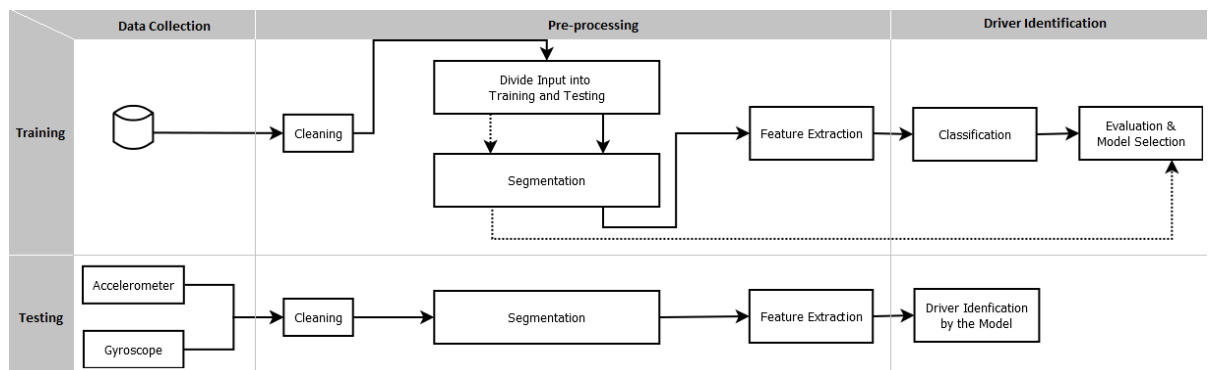

Figure 1: The overall architecture of the driver identification system

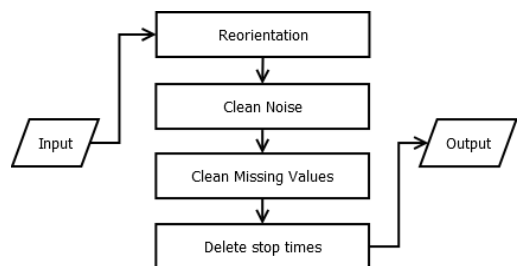

Figure 2: Data cleaning structure

In the training phase, the input data is divided into the train and test sections. The training data are used for the training and model selection and the test data are used to evaluate the model. In Fig. 1 these two sections are shown separately. Next, the data will be segmented. The test phase windows are independent of the training phase windows and there will be no overlap between the training and test data, but in both phases, the windows overlap to a certain extent. Depending on the window length and the size of overlap specified, the number of different windows can be considered for this data. For example, there is a window with a 25-minute length and a 50% overlap. The results of the different length windows will be compared in the next chapter.

In the feature extraction step, the histogram, mean, variance, difference, and correlation for each window are obtained. In extracting the histogram, the quantities are considered to be 100. This number is obtained by using the Grid Search algorithm and also to eliminate the outliers of the histogram, the data rage reduced to 95% according to the 68–95–99.7 rule. In other words, the first and fourth quarters are removed from the distribution of each window. In extracting the statistical features, if each window contains 1200 samples, the mean and variance of these 1200 samples and the difference between the sum of each window and the average of the previous window are obtained. The correlation of each window is also obtained relative to the other signals at the same time. Assuming all three-axis accelerometers and gyroscopes are used for 6 input windows at the same time, we will have 6 averages, 6 variances, 6 difference, and 15 correlation variables. Finally, the features extracted from the windows will be standardized based on the training data.

In the identification module, in the training phase different machine learning algorithms are used based on the extracted features and then with the test data, the performance of the learning algorithm is evaluated. This study attempts to test a wide range of machine learning algorithms. For this purpose, Adaboost, Gradient Boosting, Decision Tree, K-Nearest Neighbor, Support Vector Machine, Logistics, Neural Network, Naive Bayesian, and Random Forest are used. After selecting the best model for driver identification, in the test phase, only the algorithm will be used.

## 3- System Evaluation

In this study, data from the accelerometer and gyroscope sensors in the dataset [30] were used. Each of the sensors used has three axes of motion monitoring data and the input system contains six accelerometer and gyroscope sensor signals. The sampling rate in the dataset is 2 Hz. Table 3 illustrates the details of the input data before and after data cleaning.

*Table 3: Statistics of input data pre-processing*

| Driver ID | Driving duration time | Duration time of stoping | Duration time of movement | Clean movement duration time |
|---|---|---|---|---|
| 201 | 4:19:09 | 2:55:37 | 1:23:32 | 1:23:32 |
| 202 | 6:53:36 | 5:10:52 | 1:42:44 | 1:42:44 |
| 203 | 8:24:28 | 6:11:51 | 2:12:37 | 2:12:31 |
| 204 | 8:39:25 | 7:12:08 | 1:27:17 | 1:27:11 |
| 205 | 9:38:07 | 8:19:03 | 1:19:04 | 1:18:58 |
| 206 | 6:44:40 | 5:08:03 | 1:36:36 | 1:36:30 |
| 207 | 12:13:19 | 10:02:49 | 2:10:30 | 2:10:24 |
| 208 | 8:00:58 | 6:13:15 | 1:47:43 | 1:47:37 |
| 209 | 6:34:13 | 4:36:12 | 1:58:01 | 1:57:55 |
| 210 | 6:05:37 | 3:40:55 | 2:24:42 | 2:24:36 |

As statistics shows, the driving data for each driver has been reduced from 7:30 to 1:50 minutes. In the following, 70% of the input data is assigned to the training and the rest to the test, and the window segmentation and extraction steps are performed separately on the training and test data. At the segmentation stage, the length and overlap parameters of the windows in separate experiments were set to 5, 10, 15 and 30 minutes and 0%, 25%, 50%, and 75%. At the feature extraction stage, different combinations of features have been investigated in separate experiments. At the classification stage, Adaboost, Gradient Boosting, Decision Tree, K-Nearest Neighbor, Support Vector Machine, Logistics, MLP, Naive Bayesian and Random Forest are investigated. According to the experiments, the best algorithms for the identification drivers are the K-Nearest Neighbor, the Decision Tree, the Random Forest, and MLP, of which the MLP with 96% accuracy performed better than the others. Table 4 reports the accuracy of different models based on different combinations of extracted features. In all these experiments the number of drivers is 10.

*Table 4: Report statistics according to different parameters*

| Window duration (minute) | Windows overlap (%) | Features | Mean of Accuracy (%) |
|---|---|---|---|
| 15 | 75 | Histogram | 93 |
| 15 | 75 | Mean, Variance, Difference, and Correlation | 90 |
| 15 | 75 | Histogram, Mean, Variance, Difference, and Correlation | 96 |
| 15 | 75 | Histogram and Mean | 88 |
| 15 | 75 | Histogram and Variance | 88 |
| 15 | 75 | Histogram, Mean, and Variance | 88 |
| 15 | 75 | Histogram, Mean, Variance, and Difference | 82 |
| 15 | 75 | Histogram and Correlation | 88 |
| 15 | 75 | Histogram, Mean, and Correlation | 88 |
| 15 | 75 | Mean | 60 |
| 10 | 75 | Variance | 35 |
| 10 | 75 | Difference | 58 |
| 10 | 75 | Correlation | 61 |
| 10 | 75 | Mean and Variance | 76 |
| 10 | 75 | Mean and Difference | 58 |
| 10 | 75 | Mean and Correlation | 81 |
| 10 | 75 | Mean, Variance, and Difference | 70 |
| 10 | 75 | Mean, Variance, and Correlation | 81 |
| 10 | 75 | Mean, Variance, Difference, and Correlation | 83 |

Table 5 compares the results and parameters of several studies related to the present paper. These results show that the proposed system is competitive with the best practices in the field of driver identification and has performed better than previous systems in many fields.

*Table 5: Comparison of results of the proposed system with previous work in the field of driver identification*

| Ref | Sensors | Duration Time of Data Collection | Features | Number of Drivers | Noise Filter | Classifier | Accuracy (%) |
|---|---|---|---|---|---|---|---|
| Current Research | Accelerometer, Gyroscope | Avg 7:30 per driver | Histogram and Statistical Features | 10 | * | MLP, KNN, Decision Tree, Random Forest | 96 |
| [27] | Accelerometer | 10 months | Histogram | 13 | * | Neural Network | 95 |
| [26] | Accelerometer | 3 months | Statistical Features with PCA | 5 | - | - | 60-100 |
| [25] | GPS | 2 months | Statistical Features | 5 | - | Random Forest | 82.3 |
| [16] | Brake pedal position | Avg 3 hours per driver | Statistical, Descriptive and Frequent Features | 15 | - | Random Forest | 87-100 |

## 4- A plan for development

Currently, the Sepahtan system is used on intercity buses to monitor drivers online, which violates the privacy of drivers due to face imaging. The cost of installing this system is also very high. Due to the limitations of implementation in Iran, the proposed system of this article is an alternative system with the appropriate cost and high privacy protection. Of course, the implementation of this plan on all cars requires the implementation of smart insurance plans that have not yet become popular in Iran. Therefore, the authors of this paper propose to launch the system on intercity buses in the first phase.

For this purpose, mobile application software can transmission inertia sensors information through the Internet of Things to a server at the police or private transport companies. It should be noted that since the information includes date and time, it is also possible to store them offline and when the Internet is established, the information is sent to the server. This case minimizes the possible concerns about the implementation of this system in Iran.

Also, because the smartphone must be consistent, a frame will be made for the mobile so that it will be there while driving. Then, using the results of the process, the bus driver can be identified and then alerted to the authorities accordingly. Since the input of driver evaluation systems is inertial sensors information, adding a driver evaluation module in addition to identifying the driver, improves enhancing driver performance and safety. In later phases, this system can be developed into intercity trucks and then taxis and private car drivers in Iran.

## 5- Conclusion

According to the results, accelerometer and gyroscope sensors have the potential to differentiate drivers based on their behavioral characteristics with the help of feature extraction and machine learning techniques. Although primitive methods such as face recognition and fingerprinting make the driver identification more accurate, but primitive methods are more likely to be deceptive than the proposed system, because motion sensors are active throughout the movement, face recognition and fingerprinting can be used once in motion. In the case of continuous shooting, the driver's comfort is also disturbed; on the other hand, accelerometer and gyroscope sensors do not violate the privacy of the driver, so this is a better option for companies such as insurance.

Various parameters including the number of drivers, window size, size of window overlap, input signals and extracted features have been investigated. With 10 drivers, 15 minutes windows and 75% overlap, using 6 input signals (3 accelerometer signals and 3 gyro signals), histogram, and statistical features, this paper achieved 96% accuracy. Better and cheaper than many similar jobs. Therefore, it is recommended to use this system in various incentives and punitive driving applications. In the next work, different feature extraction methods can be used to measure the effect of non-statistical methods along with statistical methods. For example, deep learning [31] can be used to extract some hidden features to identify drivers, in general. Also discrete wavelet [32] can be applied to extract the features from driver signals. The comparison between these features are left to the future works. On the other hand, it is possible to develop a model under uncertainty to identify driver. In a former research an ANFIS has been used to evaluate the driving styles [33]. In the next researchers the same methodology can be used to identify the drivers. Also to improve the driver identification, it is a novel approach to evaluate the driver in the different modes. Mode detection is a standard problem that has been solved by smartphone sensors data [34]. One can combine the approach of this paper for different modes in the next studies. In the case of the different techniques of data mining on smartphone data instead of the approach of this paper, one can follow the methods in [35]. Many techniques of data mining can be used to identify the drivers in the next researches. The last but not the least, for driving identification, the overfitting makes very bad effects on the performance of system and the reliability decreases essentially. It is a scientific gap to find an appropriate regularization [36] for driving identification before releasing a commercial software.

## 6- References